\documentclass[fleqn,usenatbib]{mnras}

\usepackage{newtxtext,newtxmath}
\usepackage[T1]{fontenc}
\usepackage{graphicx}	
\usepackage{amsmath}	

\title[SNR Faraday Tomography]{On the Potential of Faraday Tomography to Identify Shock Structures in Supernova Remnants}

\author[S. Ideguchi et al.]{
Shinsuke Ideguchi,$^{1,2}$\thanks{E-mail: s.ideguchi@astro.ru.nl}
Tsuyoshi Inoue,$^{3,4}$
Takuya Akahori$^{5,6}$
and Keitaro Takahashi$^{2,5,7}$
\\
$^{1}$Department of Astrophysics/IMAPP, Radboud University, PO Box 9010, 6500 GL Nijmegen, The Netherlands\\
$^{2}$Faculty of Advanced Science and Technology, Kumamoto University, 2-39-1, Kurokami, Kumamoto 860-8555, Japan\\
$^{3}$Department of Physics, Konan University, Okamoto 8-9-1, Kobe, Japan\\
$^{4}$Department of Physics, Graduate School of Science, Nagoya University, Furo-cho, Chikusa-ku, Nagoya 464-8602, Japan\\
$^{5}$Mizusawa VLBI Observatory, National Astronomical Observatory of Japan, 2-21-1 Osawa, Mitaka, Tokyo 181-8588, Japan\\
$^{6}$Operation Division, Square Kilometre Array Observatory, Lower Withington, Macclesfield, Cheshire SK11 9FT, UK\\
$^{7}$International Research Organization for Advanced Science and Technology, Kumamoto University, 2-39-1 Kurokami, Chuo-ku, Kumamoto 860-8555, Japan
}

\date{Accepted XXX. Received YYY; in original form ZZZ}
\pubyear{2022}

\begin{document}
\label{firstpage}
\pagerange{\pageref{firstpage}--\pageref{lastpage}}
\maketitle

\begin{abstract}
Knowledge about the magnetic fields in supernova remnants (SNRs) is of paramount importance for constraining Galactic cosmic ray acceleration models. It could also indirectly provide information on the interstellar magnetic fields. In this paper, we predict the Faraday dispersion functions (FDFs) of SNRs for the first time. For this study, we use the results of three dimensional (3D) ideal magnetohydrodynamic (MHD) simulations of SNRs expanding into a weak, regular magnetic field. We present the intrinsic FDFs of the shocked region of SNRs for different viewing angles. We find that the FDFs are generally Faraday complex, which implies that conventional rotation measure study is not sufficient to obtain the information on the magnetic fields in the shocked region and Faraday tomography is necessary. We also show that the FDF allows to derive the physical-depth distribution of polarization intensity when the line of sight is parallel to the initial magnetic field orientation. Furthermore, we demonstrate that the location of contact discontinuity can be identified from the radial profile of the width of the FDF with the accuracy of 0.1-0.2 pc.
\end{abstract}




\section{Introduction}
\label{sec:1}

Diffusive shock acceleration \citep[DSA; see e.g.,][for a review]{blandford_eichler_1987} is broadly accepted as the main acceleration process of cosmic-rays in supernova remnants (SNRs) and the intra-cluster medium (ICM). The detailed physics and efficiency of this acceleration mechanism is, however, a longstanding question.

In the DSA theory, particle acceleration takes place around shock structures such as forward shocks (FS), and reverse shocks (RS). \cite{decourchelle_etal_2000} found that efficient particle acceleration at the FS (RS) shrinks the gap between the contact discontinuity (CD) and the FS (RS) due to a higher compressibility of the plasma than in the adiabatic case. X-ray observations have reported such shrinking of the gap between the FS and the CD in SN 1006 \citep{Cassam-Chenai_etal_2008} and Tycho's SNR \citep{warren_etal_2005}.
Here, \cite{Cassam-Chenai_etal_2008} determined the location of the CD from the low-energy X-ray band (0.5 - 0.8 keV), assuming that the oxygen emission seen in the band mostly comes from the shocked ejecta. But an alternative possibility is that the emission is associated with the shocked ambient medium \citep{yamaguchi_etal_2008}. 
Therefore, a reliable way to identify the location of shock structures has also important implications for the question of the particle acceleration.

The structure of the magnetic field around the shock plays an essential role in DSA efficiency. The standard scenario assumes indeed that particles are repeatedly scattered by magnetic fluctuations back and forth across the shock, with energy gain at each crossing. It is also useful to understand the origin and evolution of magnetic fields there \citep{reynolds_etal_2012}. The connections between the magnetic fields of SNRs and that of the ambient interstellar medium (ISM) have been studied by means of observations \citep[e.g.][]{purcell_etal_2015, sun_etal_2015} and numerical simulations \citep[e.g.][]{petruk_etal_2009,schneiter_etal_2010,bocchino_etal_2011,schneiter_etal_2015,west_etal_2016, west_etal_2017a,west_etal_2017b,winner_etal_2020}. Such a connection implies a potential correlation between the magnetic fields of SNRs and the global magnetic field of the Milky Way. Hence, the detailed study of the SNR magnetic fields is essential not only to understand the SNR itself but also to figure out effects of the ambient environment on the SNR.

Radio spectro-polarimetric observation of synchrotron radiation is a unique probe of the strength and orientation of SNR magnetic fields \citep[see e.g.,][for a review]{dubner_giacani_2015}. Such observations provide information about the Faraday rotation: the plane of the linear polarization rotates in proportion to the squared wavelength ($\lambda^2$) and the Faraday rotation measure (RM) such that
\begin{equation}
\Delta\chi=\chi-\chi_0={\rm RM}\lambda^2,
\label{eq:rm}
\end{equation}
where $\chi$ and $\chi_0$ are the observed and intrinsic polarization angles, respectively. The RM, which is estimated by the linear fitting of $\chi$ vs. $\lambda^2$ plot, is expressed as
\begin{equation}\label{eq:RM}
{\rm RM}=0.812\int_{\rm source}^{\rm observer}n_{\rm e}(r')B_{\parallel}(r') dr'~[{\rm rad~m^{-2}}],
\end{equation}
where $n_{\rm e}$ is the electron number density in units of cm$^{-3}$, $B_{\parallel}$ is the line of sight (LOS) component of the magnetic field in $\mu$G, and $r$ is the physical distance in pc. The RM allows us to investigate the properties of magnetized plasma along the LOS.

This technique has been applied to several SNRs to study themselves and their foreground. \cite{kothes_brown_2009} showed that the compressed regular field can be recognized via an RM gradient along both shells (see also Appendix \ref{sec:app1}), and that only two SNRs, DA530 and the southern shell of G182.4+4.3, are roughly consistent with this scenario. CTA1 (G119.5+10.2) shows different RM signs in the two shells which is possibly due to a Faraday screen in front of the SNR covering one shell \citep{sun_etal_2011}. \cite{reynoso_etal_2013} performed an RM study on the SN 1006 and found that the RM value in the NE and SW lobes of the shell is 12 rad m$^{-2}\pm$ 20 rad m$^{-2}$ and that the orientation of magnetic field vectors appear to be mostly radial. They also found that SN1006 does not show any systematic RM pattern. Excess RM is detected near the northern rim of the Gum nebula but with a large dispersion, and the compression factor is found to be too high for adiabatic compression \citep{purcell_etal_2015}. The RMs observed towards the North Polar Spur most probably occur in the foreground \citep{sun_etal_2015}. The shells of G181.1+9.5 reveal different RM signs and no RM gradient, indicating RMs from the foreground or a precursor wind \citep{kothes_etal_2017}. The large RM values observed in parts of G57.2+0.8 and the lack of Faraday depolarization also indicate RMs from the foreground \citep{kothes_etal_2018}. The shells of G1.9+0.3 reveal RM gradients but strongly different average RMs, possibly due to a progenitor wind \citep{luken_etal_2020}. A jump in the RM observed along shell C of CTB80 without corresponding Faraday depolarization indicates a foreground effect \citep{li_etal_2020}.

Broad-band radio polarimetry, using the Square Kilometre Array (SKA) and its pathfinders/precursors such as LOFAR, ASKAP, MeerKAT, MWA and HERA, will certainly revolutionize cosmic magnetism science through the Faraday tomography technique \citep{brentjens_debruyn_2005, heald_2009, akahori_etal_2018}. This technique enables us to estimate the Faraday dispersion function (FDF, or the Faraday spectrum), $F(\phi)$, from the polarization spectrum, $P(\lambda^2)$, by the following equation,
\begin{equation}\label{eq:PI}
P(\lambda^2)=\int_{-\infty}^{+\infty}F(\phi)e^{2i\phi\lambda^2} d\phi.
\end{equation}
Here, $\phi$ is called the Faraday depth, it is equivalent to the RM when $F(\phi)$ is expressed as a single delta function \citep{sokoloff_etal_1998}, and is defined as
\begin{equation}\label{eq:phi}
\phi(r)=0.812\int_{r}^{0}n_{\rm e}(r')B_{\parallel}(r') dr'~[{\rm rad~m^{-2}}].
\end{equation}
The FDF indicates the synchrotron polarization intensity as a function of $\phi$. Since broad-band data is crucial to obtain high-quality FDF, observations of new generation radio telescopes mentioned above are essential for this technique. There have been few works applying the Faraday tomography technique to SNR observations. \cite{ng_etal_2012} concluded that the uncertainty in $\phi$ space was too large to be useful for studying SNRs in detail from a single LOS, while \cite{harvey-smith_etal_2010} made a Faraday depth cube by applying the technique to many LOSs. It provides the overall RM feature such as anti-symmetric pattern in the SNR G296.5+10.0, possibly shaped by the magnetic field of a progenitor wind.
 
The traditional RM study using Eq. \ref{eq:rm} is valid only for Faraday simple situation where there is a single Faraday-thin component with a non-emitting foreground. On the other hand, the Faraday tomography technique is applicable even when the Faraday structure is complex which generally provides a non-linear relation between $\chi$ and $\lambda^2$. Moreover, the technique provides the LOS distribution of polarized sources which allows to study e.g. the polarization structure of spatially unresolved active galactic nuclei observations \citep{osullivan_etal_2012} and the strength and scale of the turbulent magnetic field in the interstellar medium \citep{ideguchi_etal_2017}. Although the near-future broader-band data would provide higher resolution in $\phi$ space and thus the detailed Faraday structures of SNRs, we have yet to understand how to extract physical information from them. Thus, it is necessary to predict the shape of the SNR FDFs and to interpret them. 

In this paper, we calculate the FDFs of LOSs going through the shocked regions of SNRs for the first time using the results of 3D ideal MHD simulations. The aim of this paper is to investigate the physical information imprinted in the FDF. For this purpose, in this trial work, we make several simple assumptions. First, we consider a simple model in which the SNR expands into an ambient ISM of constant gas density and regular magnetic field, and the turbulent fields in the ambient ISM is not incorporated. We note that such turbulent fields can play an important role for the evolution of radio SNRs: they are strongly amplified from the initial turbulent fields and give rise to efficient acceleration of cosmic rays \citep{inoue_etal_2009,inoue_etal_2012,inoue_etal_2013,caprioli_spitkovsky_2014a,caprioli_spitkovsky_2014b,delValle_etal_2016,xu_lazarian_2017,winner_etal_2020}. We also ignore the upstream field amplification caused by the cosmic-ray streaming instability \citep{bell_etal_2004}. Such strong magnetic fields that have been produced in the upstream region may be efficiently damped in the downstream region \citep{pohl_etal_2005}. Since we only discuss the downstream region, this amplification may not affect the results of this paper, while a strong shock may induce strong turbulence in the downstream region even if the external field in the ISM is regular \citep{pais_etal_2020}. We also note that the alternative mechanism of the turbulent dynamo can increase the downstream magnetic field \citep{giacalone_jokipli_2007,ji_etal_2016}. Such processes are not considered in this work to keep the interpretation of the results simple. Hence, our model is similar to previous analytical works such as \cite{kothes_brown_2009} and \cite{west_etal_2016} and thus, it may be applicable to, e.g. the clearly-defined shell-type SNRs listed in \cite{west_etal_2016}. On the other hand, our work relies on an MHD simulation which solves the expansion phase self-consistently and includes other effects at play in realistic settings such as the Rayleigh-Taylor instability and the associated magnetic field amplification (see Section \ref{sec:2.1}).

In the simulation we use, the explosion occurs within a weak, regular magnetic field and the synchrotron emission from the SNR has therefore a nontrivial radial profile. Thus, the global shape of the SNR depends on the viewing angle. Such dependency of the different aspect angles was studied using a simplified analytic model by \cite{kothes_brown_2009} (see also Appendix \ref{sec:app1}). The dependency of the aspect angles would also be useful to estimate the orientation of the global field of the Milky Way around target SNRs \citep[See also][]{west_etal_2016}.

The plan of our paper is as follows. In Section \ref{sec:2}, we describe the SNR simulation and the way to calculate its FDF. The results are shown in Section \ref{sec:3}. We demonstrate how we can identify the contact discontinuity from the FDF in Section \ref{sec:4}, and the discussion follows in Section \ref{sec:5}. We conclude in Section \ref{sec:6}.

\section{Model and Calculation}
\label{sec:2}

\subsection{Supernova Remnant Simulation}
\label{sec:2.1}

In this paper we use the result of an ideal 3D MHD simulation. The simulation employs a second-order Godunov-type scheme with approximate Riemann solver developed by \cite{sano_etal_1999}. The divergence free (${\rm div} B=0$) condition is ensured through the use of the constrained transport technique. Thus, we do not have to pay attention to the artificial effect of ${\rm div}B=0$ violation. A cubic numerical domain of the side length 10 pc is resolved with 512$^3$ cells (resolution $\simeq 0.02$ pc\footnote{As a reference, the resolution of 0.02 pc corresponds to $\sim$ 1'' for Tycho's SNR assuming that the distances to it is 4 kpc \citep{hayato_etal_2010}.}) which is filled by a uniform fully ionized plasma with number density $n=0.5$ cm$^{-3}$ and temperature $T=8,000$ K. We assume a mean mass of gas particles $m=1.27$ m$_{\rm proton}$ (9\% of the gas is helium) and an adiabatic index of $\gamma =5/3$. To mimic supernova explosion, we initially set a uniformly expanding ejecta of radius $R_{\rm ej}=1.5$ pc at the center of the numerical domain. The total kinetic energy of the ejecta is set to $10^{51}$ ergs, the mass to 3.0 M$_{\rm sun}$ and the thermal energy to be 5\% of the kinetic energy. As seed fluctuations to the Rayleigh-Taylor instability, we put density fluctuations to the ejecta with the standard deviation of 57\% to the mean ejecta density. The scales of the fluctuations are in the range of $R_{\rm ej}^{-1}\le |k|/2\pi\le 4\,R_{\rm ej}^{-1}$. The initial magnetic field is regular and along the $y$-direction with a magnitude of 3 $\mu$G. We use the simulation of a young SNR at $t=800$ years after the explosion.

\begin{figure}
	\includegraphics[width=\columnwidth]{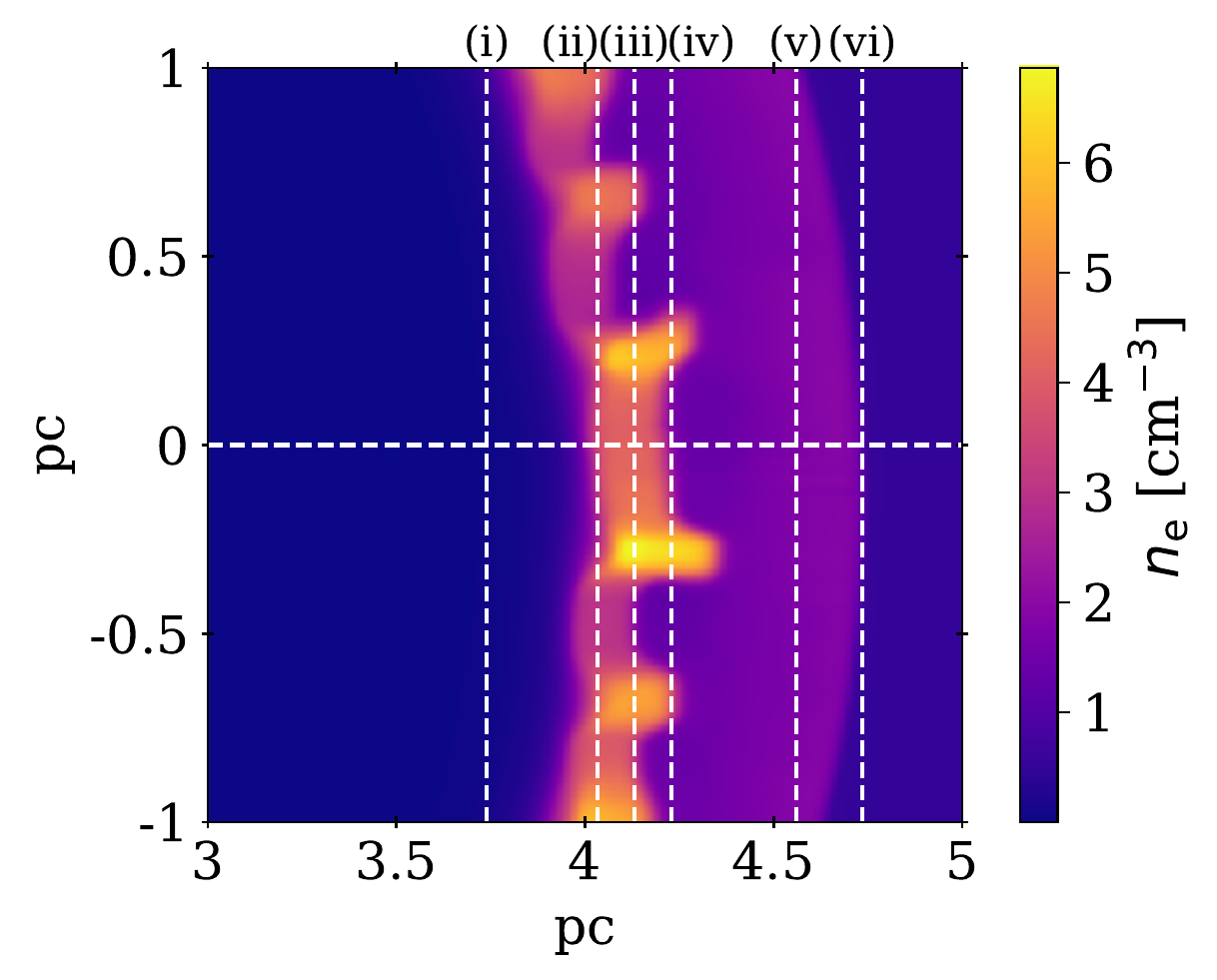}
	\includegraphics[width=\columnwidth]{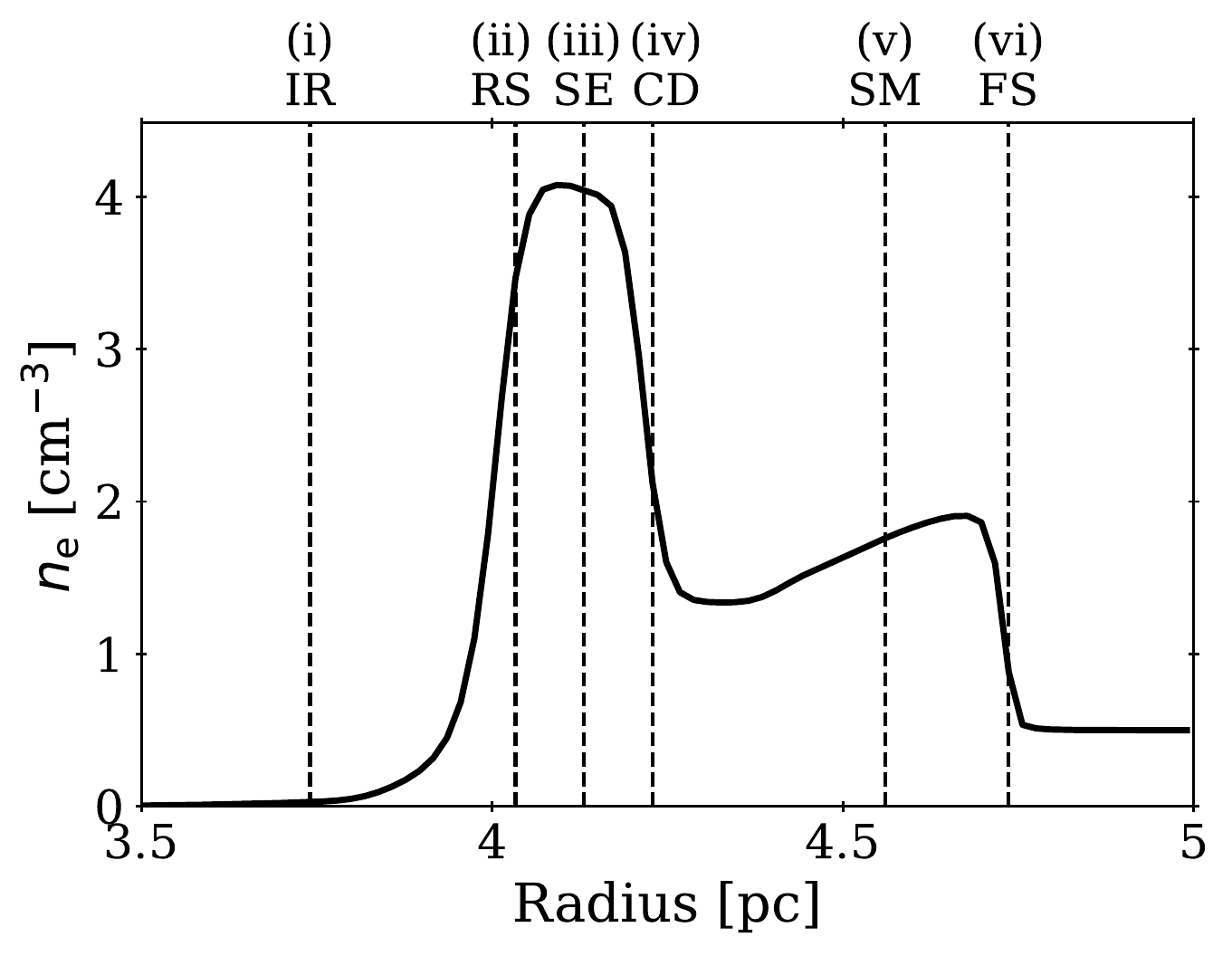}
    \caption{Top: Slice map of $n_{\rm e}$ around the rim of the SNR through the SNR center. Bottom: Radial profile of $n_{\rm e}$ along the white dashed horizontal line in the top panel. The dashed vertical lines in the bottom panel indicate the six characteristic structures, and their numbers and labels are put at the top of the panel: inner region (IR), reverse shock (RS), shocked ejecta (SE), contact discontinuity, shocked ISM (SM) and forward shock (FS). The corresponding structures are shown by dashed vertical lines and the same numbers in the top panel.
    }
    \label{fig:rho}
\end{figure}

The top panel of Figure \ref{fig:rho} shows the two-dimensional distribution of the number density, $n_{\rm e}$. A rim region of the cross section sliced at the center of the SNR sphere is shown. We can see the turbulent pattern at the rim, the consequence of the Rayleigh-Taylor instability, induced by the interaction of the compressed, dense ejecta and the light ambient ISM. The one-dimensional radial profile of $n_{\rm e}$ around the rim is shown in the bottom panel of Figure \ref{fig:rho}. The shocked region can be clearly seen in the plot. We put labels at the locations of six characteristic structures for the later analysis: (i) inner region (IR), (ii) reverse shock (RS), (iii) shocked ejecta (SE), (iv) contact discontinuity (CD), (v) shocked ISM (SM) and (vi) forward shock (FS). We note that a single LOS can pass through more than one bubble of the turbulent pattern due to the Rayleigh-Taylor instability, and/or the two (near and far) shocked regions.

\begin{figure}
	\includegraphics[width=\columnwidth]{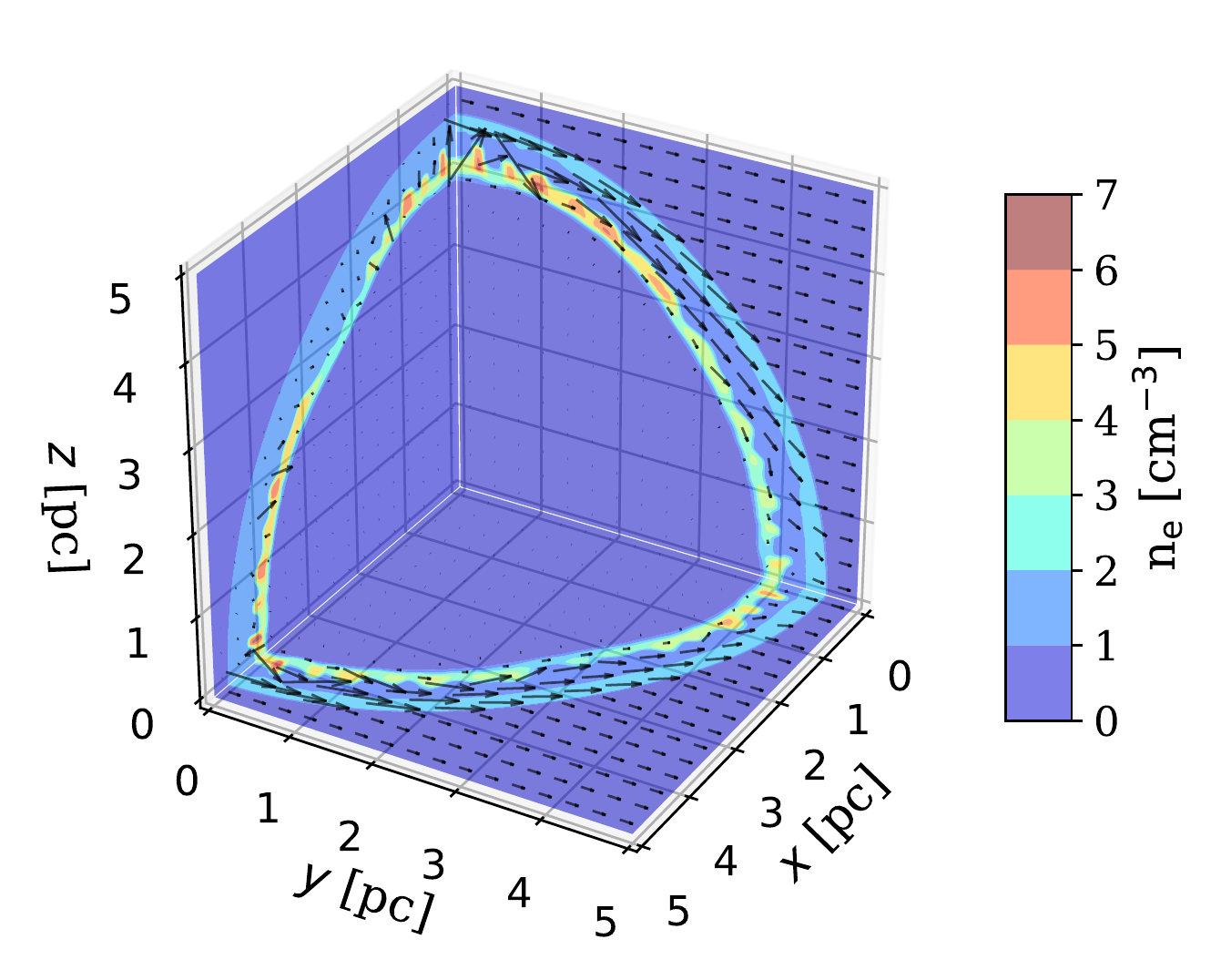}
	\includegraphics[width=\columnwidth]{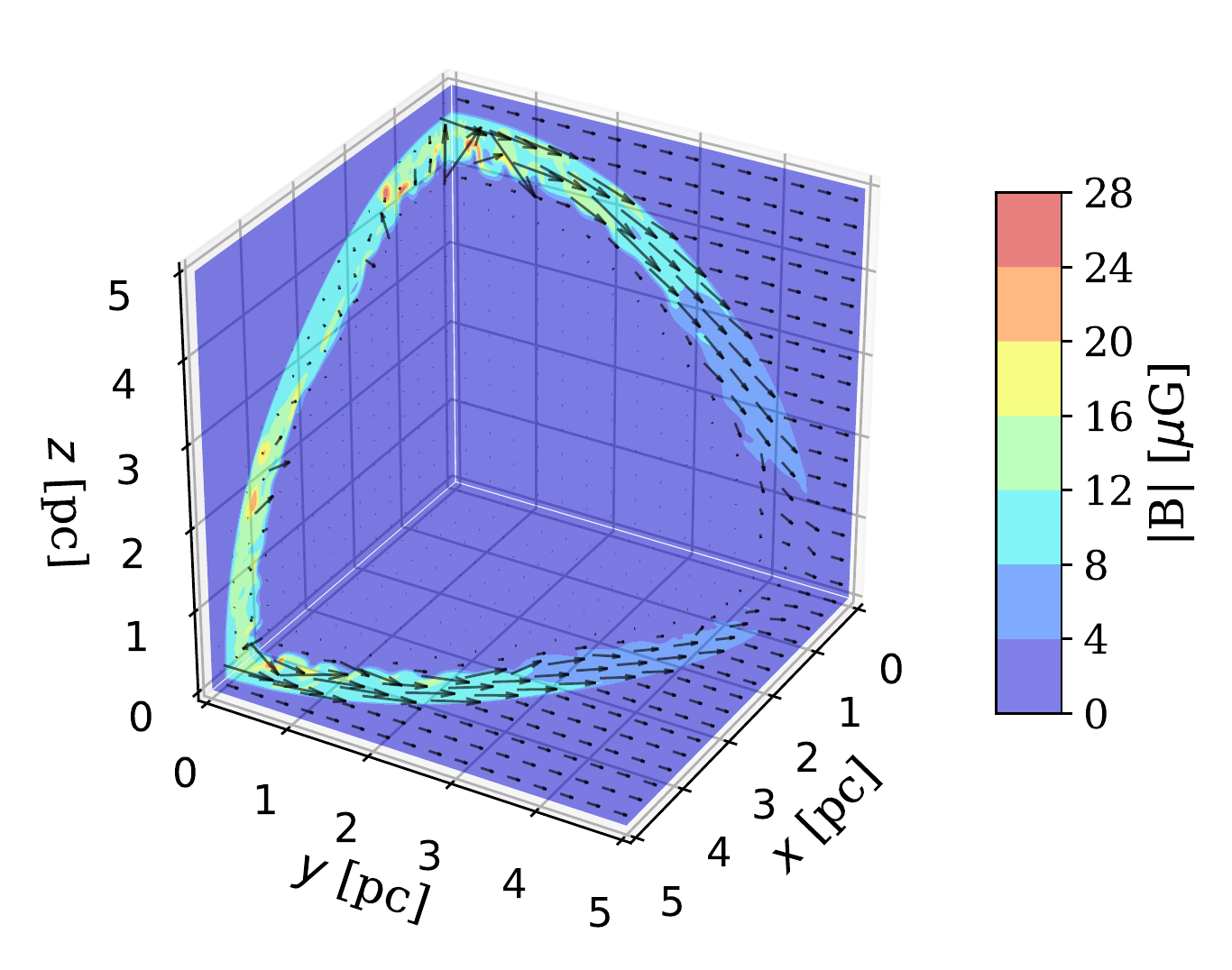}
    \caption{Slice maps of $n_{\rm e}$ (top) and $|B|$ (bottom) in the first octant of the simulation box. The black arrows in the two images represent the magnetic field vectors parallel to each plane and their length indicates the amplitude of the field. The same vector plot is overlaid in the two panels for comparison.}
    \label{fig:mag_3d}
\end{figure}

Figure \ref{fig:mag_3d} is the slice maps of $n_{\rm e}$ (top) and the magnetic field strength $|B|=\sqrt{B_x^2+B_y^2+B_z^2}$ (bottom) passing through the center of the numerical domain and through the first quadrant planes perpendicular to the $x$, $y$ and $z$-axes, which are seen as the first-octant, three-dimensional view. The over-plotted black arrows represent the field vectors parallel to the each plane. The vectors are plotted every 16 computational grids to make them easily visible. The length of the arrows indicates the amplitude of the field which is 3 $\mu$G outside of the shocked region. We can see that the magnetic field, which is initially regular and along the $y$-direction, is swept up by the SNR shell and is oriented along the surface of the SNR: Magnetic fields within the shell are mostly perpendicular to the radial direction. The field is mostly amplified at the shocked region, (ii) RS - (vi) FS, except for the intersections of the shell and the $y$-axis (i.e. the polar caps), where the ambient fields are quasi-parallel to the shock normal and are barely affected by the compression of the gas. This amplification is not only due to the adiabatic compression at the shock, but also due to the stretching of the magnetic field line by the vortex caused by the Rayleigh-Taylor instability \citep{jun_etal_1995}. In fact, in the simulation, the largest magnetic field strength is 50.5 $\mu$G (the amplification factor of $\sim$ 17), while the gas density at the same pixel is 1.57 cm$^{-3}$ (the factor of $\sim$ 3). Recent X-ray observations show that the magnetic field is amplified by a much larger factor, $\gtrsim$ 100 \citep{uchiyama_etal_2007,reynolds_etal_2012,dubner_giacani_2015}, which is most likely due to the upstream density fluctuation in the ISM \citep{inoue_etal_2009,inoue_etal_2012,inoue_etal_2013}. We ignore such turbulence in this paper and thus, do not take this additional amplification into account.

\subsection{Viewing angle}
\label{sec:2.2}

\begin{figure*}
	\includegraphics[width=0.8\textwidth]{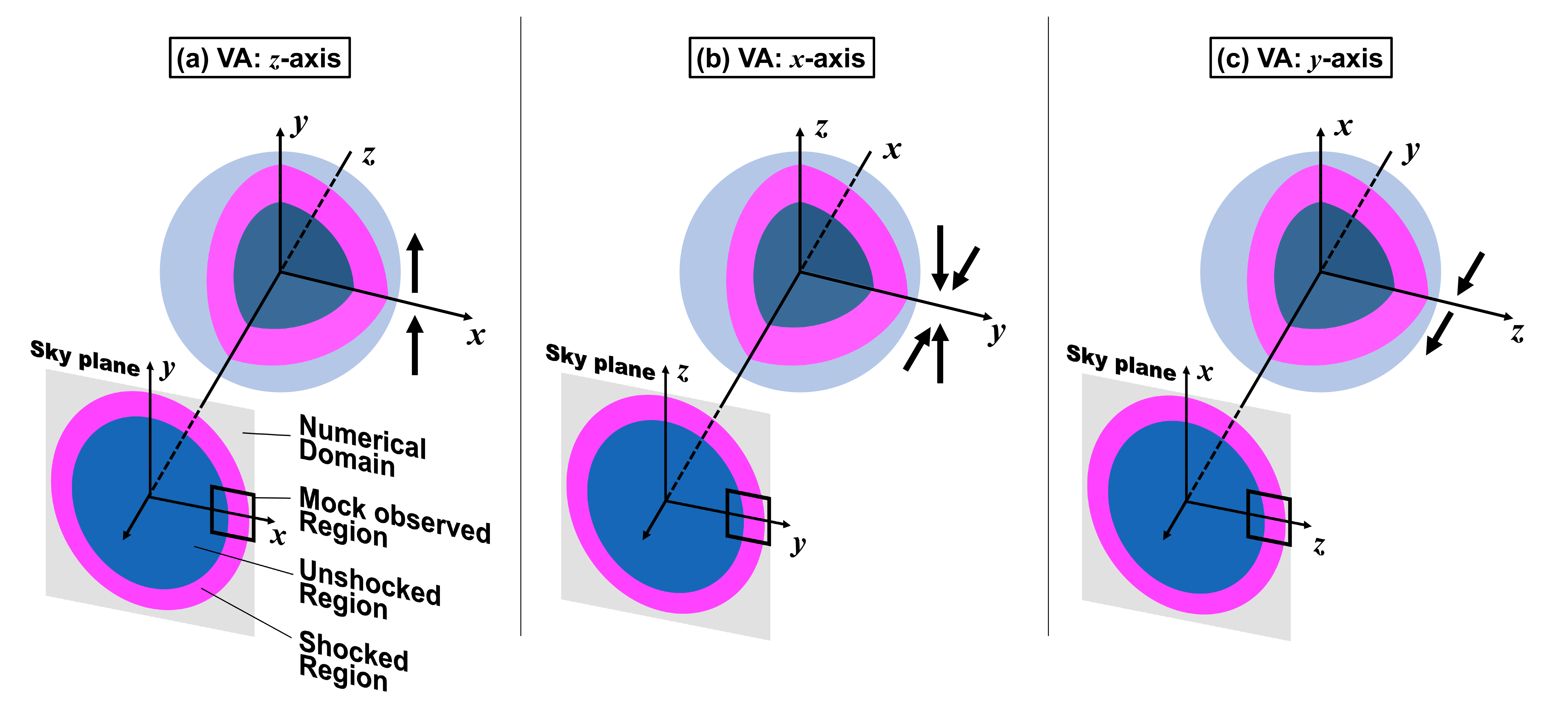}
    \caption{Schematic pictures of SNR shell and projected images along the three viewing angles (VAs). The light blue spherical shell with partial cross-sections shows the SNR shell. The VA is along the (a) $z$, (b) $x$ and (c) $y$-axes from left to right and the projected sky corresponds to the $x$-$y$, $y$-$z$ and $z$-$x$ planes, respectively. In the sky plane images, colors indicate numerical domain (grey), shocked region (magenta) and unshocked region (dark blue), and the thick lined squares show mock observed region (2$\times$2 pc$^2$) for this study. We also put thick arrows in the right side of the three dimensional plots which show the characteristic magnetic field configuration at the observed region.}
    \label{fig:view}
\end{figure*}

As mentioned in Section \ref{sec:2.1}, the magnetic fields in the simulated SNR basically point toward $y$-direction and along the shell. Hence, the magnetic field configuration can be different depending on the viewing angle and location of the SNR through which the LOS passes. In the shell of the simulated SNR used in this paper, there are essentially three independent magnetic field configurations: The field is basically
\begin{description}
    \item[(a)] perpendicular to the LOS.
    \item[(b)] parallel to the LOS and there is the field reversal.
    \item[(c)] parallel to the LOS and points all the way along the same direction.
\end{description}
In this paper, we only focus on these three magnetic field configurations since any magnetic field configuration along the LOS can be explained by a combination of them. Specifically, we choose the viewing angle along the $z$-, $x$- and $y$-axes and look at 3 to 5 pc in horizontal axis and -1 to 1 pc in vertical axis (2$\times$2 pc$^2$). The schematic picture of the three viewing angles are shown in Figure \ref{fig:view} (see the caption for detail). With this setup, the situations where the viewing angle are along the $z$-, $x$- and $y$-axes correspond to (a), (b) and (c) cases mentioned above, respectively.

\subsection{Calculation of FDF}
\label{sec:2.3}

The FDF for a given LOS through the simulated SNR can be calculated with the Faraday depth and the synchrotron emissivity both as functions of the physical depth. The Faraday depth, $\phi$, can be calculated using Equation~(\ref{eq:phi}). Specifically, $\phi$ is calculated in every cells using $n_{\rm e}$, $B_{\parallel}$, and the cell size, and the values are integrated through each LOS. Here, we assume that the polarization from a cell experiences the Faraday rotation due to the field within its own cell. 

The synchrotron emissivity depends on the cosmic-ray electron (CRe) density, $\rho_{\rm CRe}$, the perpendicular component of the field along the LOS, $B_{\perp}$, the radio frequency, $\nu$, and the index of the energy spectrum of cosmic-ray electron, $p$:
\begin{equation}
\varepsilon_{\rm sync}\propto \rho_{\rm CRe}~B_{\perp}^{(1+p)/2}~\nu^{(1-p)/2}.
\end{equation}
The polarization fraction of synchrotron radiation, $\Pi$, can be expressed in terms of $p$ such that $\Pi=(3p+3)/(3p+7)$ \citep{rybicki_lightman_1986}. In this study, we essentially focus on the ``shape'' of the intrinsic FDFs of the simulated SNR and only consider the relative intensity of the radiation. We also ignore the frequency dependence \footnote{See Section 3 of \cite{brentjens_debruyn_2005} for the effect of the frequency dependence on Faraday tomography.}. Thus, values of emissivity do not have the absolute meaning in this study, and we present results using normalized values so that one can compare the relative intensities. For the CRe density, $\rho_{\rm CRe}$, it is simply assumed to be constant. The spectral index is assumed to be $p=2$ in line with SNR radio observations \citep[see Figure 1 of][which is made from the SNR catalog in \citealt{green_2009}]{reynolds_etal_2012}. We also assume that the spectral index is constant in every computational cells. Hence, the synchrotron radiation can simply be calculated as
\begin{equation}\label{eq:pol}
\varepsilon_{\rm sync}\propto B_{\perp}^{1.5}.
\end{equation}
Note that, as $p$ is constant, the polarization fraction of the synchrotron radiation is constant in each cell. Thus, we can treat the synchrotron emissivity as the polarization intensity because we only consider the relative intensity of the radiation, as mentioned above. The linear polarization is calculated so that its orientation is perpendicular to the magnetic field vector. Finally, we ignore the emission and the Faraday rotation outside of the SNR to only focus on the FDFs of the SNR itself. For this purpose, we set the amplitude of the magnetic field outside the shocked region, (vi) FS, to 0.

\begin{figure*}
	\includegraphics[width=0.8\textwidth]{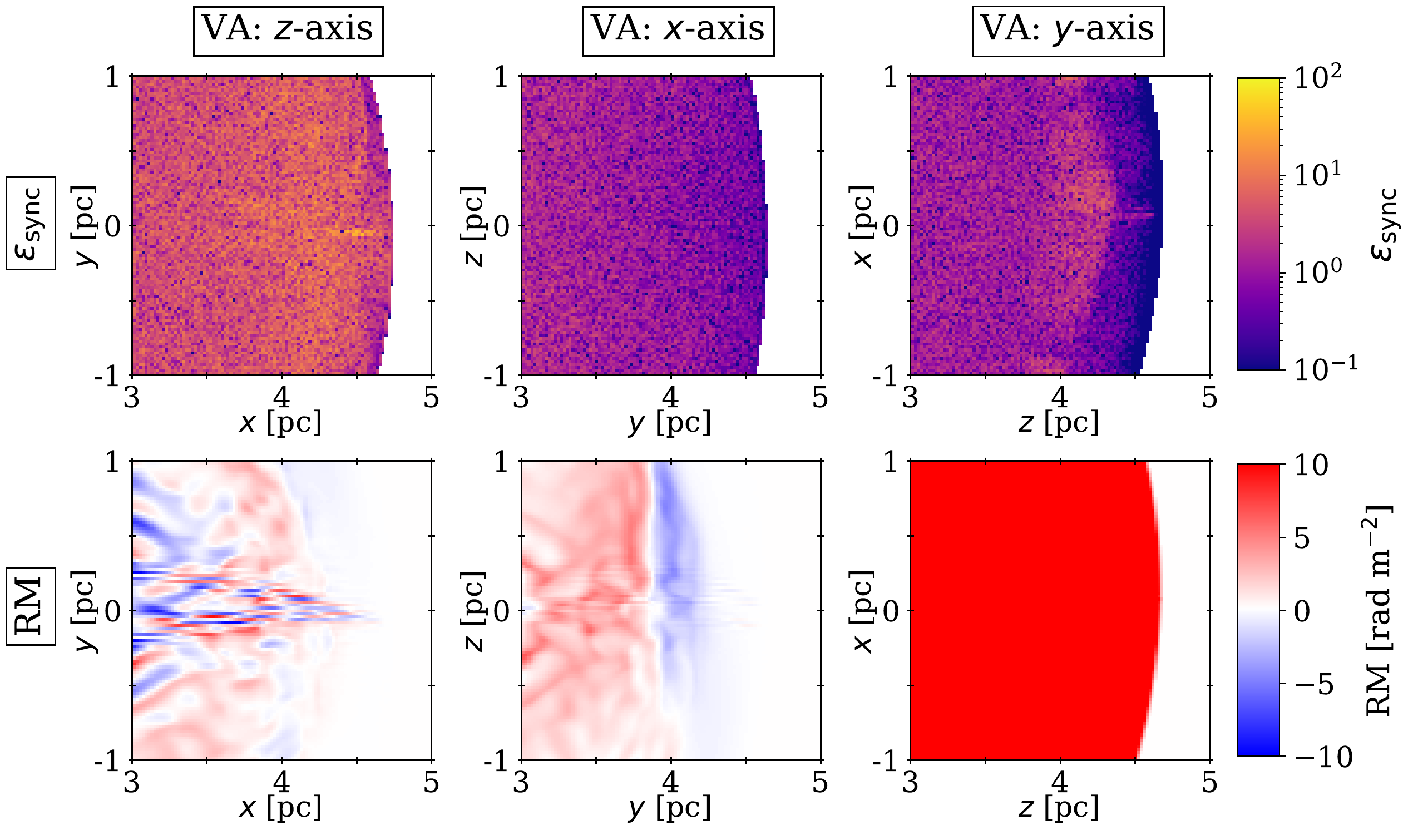}
    \caption{Projected maps of synchrotron polarization (top) and RM (bottom) after including the polarization radiative transfer. Three viewing angles (VAs) are shown: VA is parallel to the $z$- (left), $x$- (middle) and $y$-axes (right). The color of the RM map for ``VA:$y$-axis'' case (bottom right panel) is scaled out because we use the same color scale for all the three cases and the scale is adjusted to clearly see the RM structures in the other two cases. Note that the RM fluctuations are attributed to fluctuations in $n_{\rm e}$ and $B_\parallel$ due to the Rayleigh-Taylor instability. We also note that the maximum RM value for ``VA:$y$-axis'' case is $\sim$ 80 rad m$^{-2}$ (see also Section \ref{sec:3.1}, Figure \ref{fig:fdf3}, and Figure \ref{fig:rad_dep}).}
    \label{fig:sync_map}
\end{figure*}

In Figure \ref{fig:sync_map}, we show the projected synchrotron maps at 1.4 GHz (top panels) and the RM maps (bottom panels) of the rims from the three viewing angles (see also Appendix \ref{sec:app1} for the detailed look at the simulation results). In the calculation of the synchrotron maps, the polarization radiative transfer and Faraday depolarization at 1.4 GHz are included. Here, the RM maps show the integration of $n_{\rm e}B_\parallel$ along the whole LOS and thus, they correspond to RM maps by using background sources of the SNR. The maps correspond to the regions delimited by the thick lined squares in Figure \ref{fig:view}. Keeping Figure \ref{fig:mag_3d} and \ref{fig:view} in mind, one can briefly interpret the results as follows. When the viewing angle is along the $z$-axis (left column), the perpendicular component of the magnetic field is relatively large and the emission is strong. On the contrary, the parallel component is very small which results in small RM values. Along the $x$-axis (middle column), the perpendicular component of the field is small since the ambient field is quasi-parallel to the shock normal and hence, barely amplified, as mentioned in Section \ref{sec:2.1}. Thus the emission is small. The RM values are also small, as in the previous case, and additionally, it changes its direction between the near and far side of the SNR along the LOS. When the viewing angle is along the $y$-axis (right column), the magnetic field approximately points toward us which results in small emission and large RM values. These basic features of the amount of emission and RM values depending on the viewing angle are summarized in Table \ref{tab:mag_conf}.

Finally, the FDF corresponds to the histogram of the emission as a function of the Faraday depth. Note that the depolarization can occur in each Faraday depth bin when the angles of polarization from more than one cell are different.

\begin{table}
 \caption{Amount of emission ($\varepsilon_{\rm sync}$) and RM depending on the viewing angles (VAs)}
 \label{tab:mag_conf}
 \begin{tabular}{cc|cl}
  \hline
  &VA & $\varepsilon_{\rm sync}$ & $\phi$ or RM\\
  \hline
  (a) & $z$ & large & small\\[2pt]
  (b) & $x$ & small & small (field reversal)\\[2pt]
  (c) & $y$ & small & large\\[2pt]
  \hline
 \end{tabular}
\end{table}

\section{FDF of Simulated SNR}
\label{sec:3}

\subsection{One Dimensional FDF}
\label{sec:3.1}

\begin{figure*}
	\includegraphics[width=0.8\textwidth]{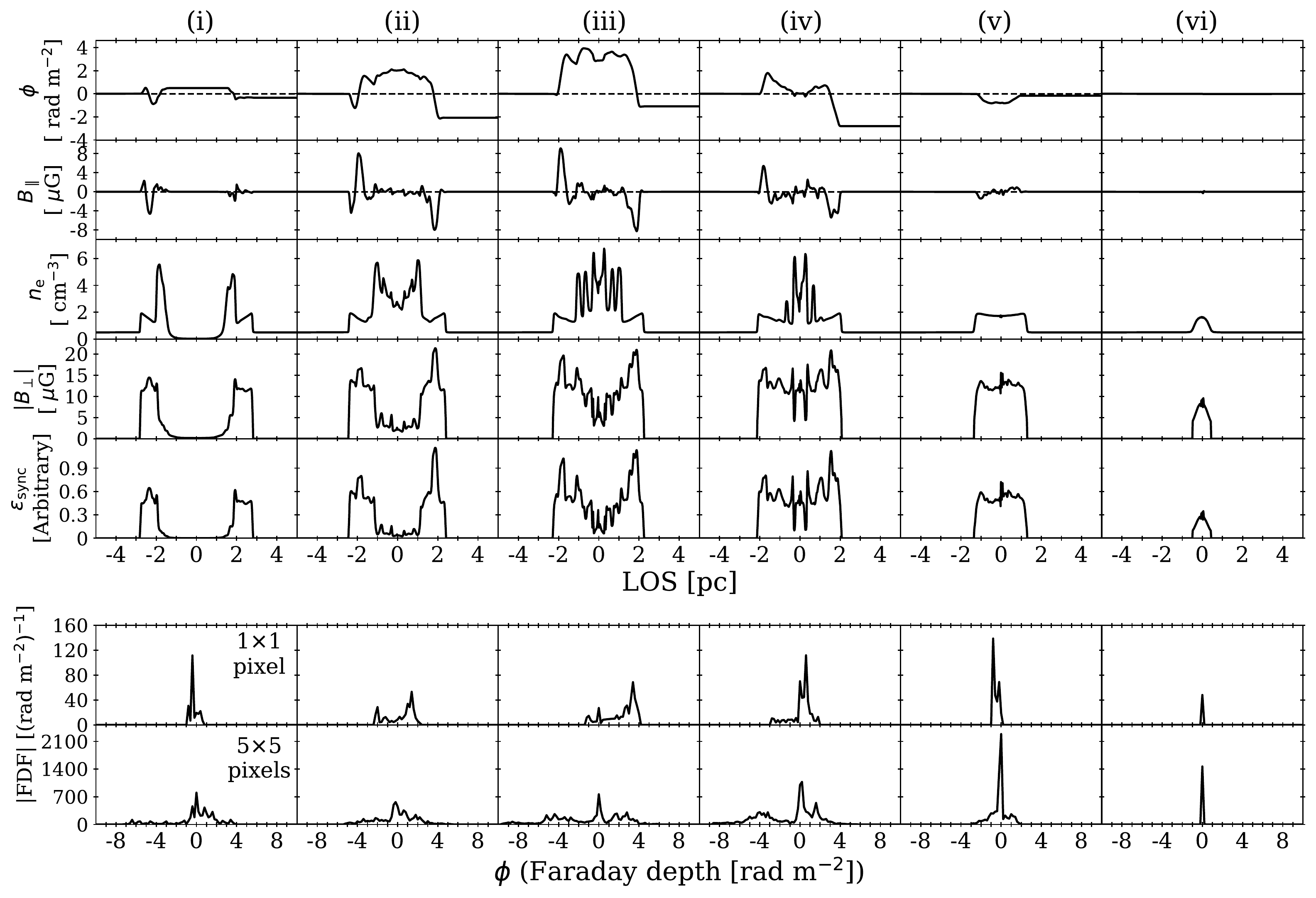}
    \caption{Results of viewing angle along the $z$-axis. LOS distributions of physical quantities $B_\parallel$, $n_{\rm e}$ and $|B_\perp|$ in second, third and fourth rows, respectively, and $\phi$ (first rows) and $\varepsilon_{\rm sync}$ (fifth rows) calculated from the above physical quantities. The bottom two panels show the resultant FDFs with 1$\times$1 and 5$\times$5 pixels on the image plane. Columns correspond to the six characteristic structures within the shock region (see also Figure \ref{fig:rho}).}
    \label{fig:fdf1}
\end{figure*}

\begin{figure*}
	\includegraphics[width=0.8\textwidth]{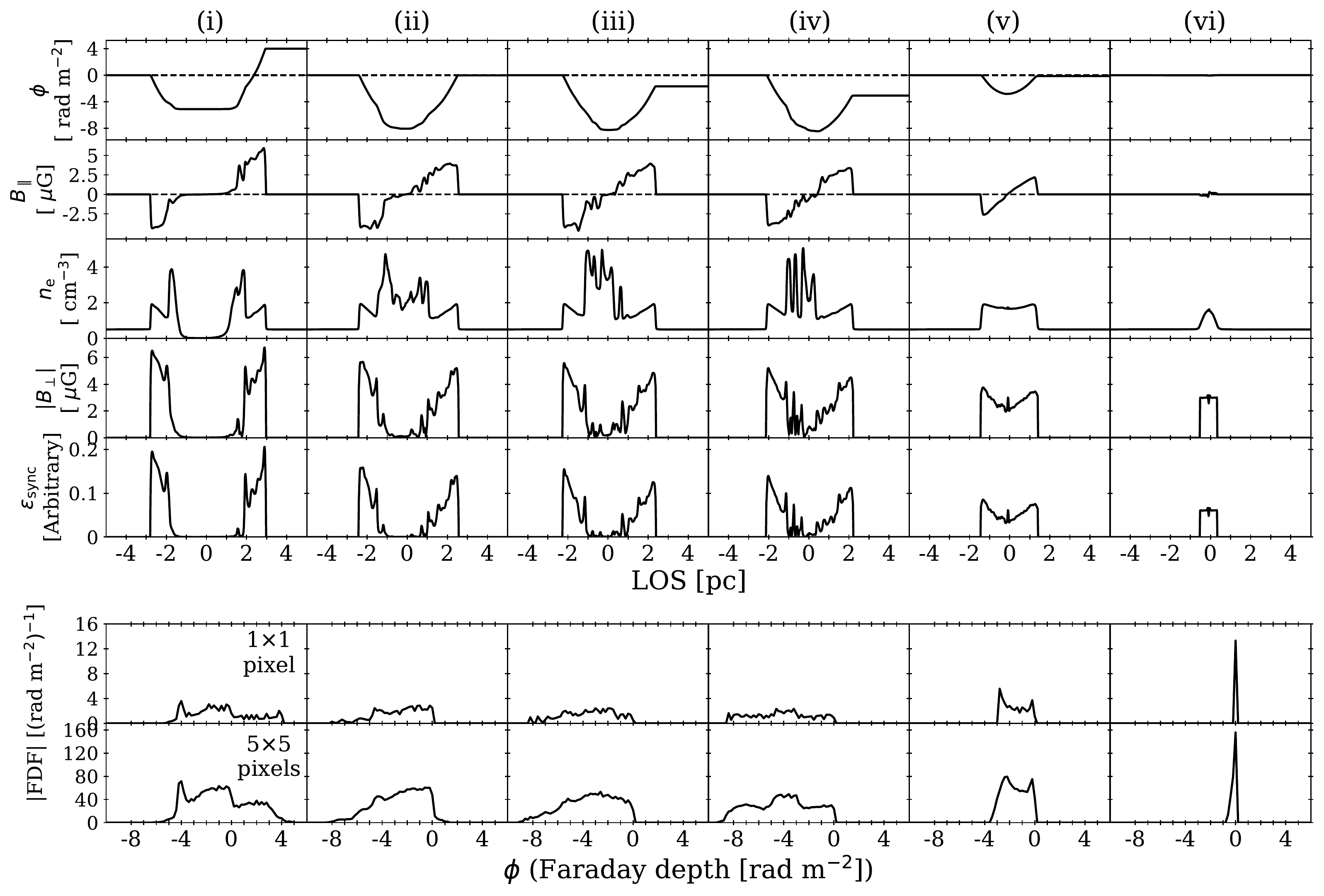}
    \caption{Same as Figure \ref{fig:fdf1}, when the viewing angle is along the $x$-axis.}
    \label{fig:fdf2}
\end{figure*}

\begin{figure*}
	\includegraphics[width=0.8\textwidth]{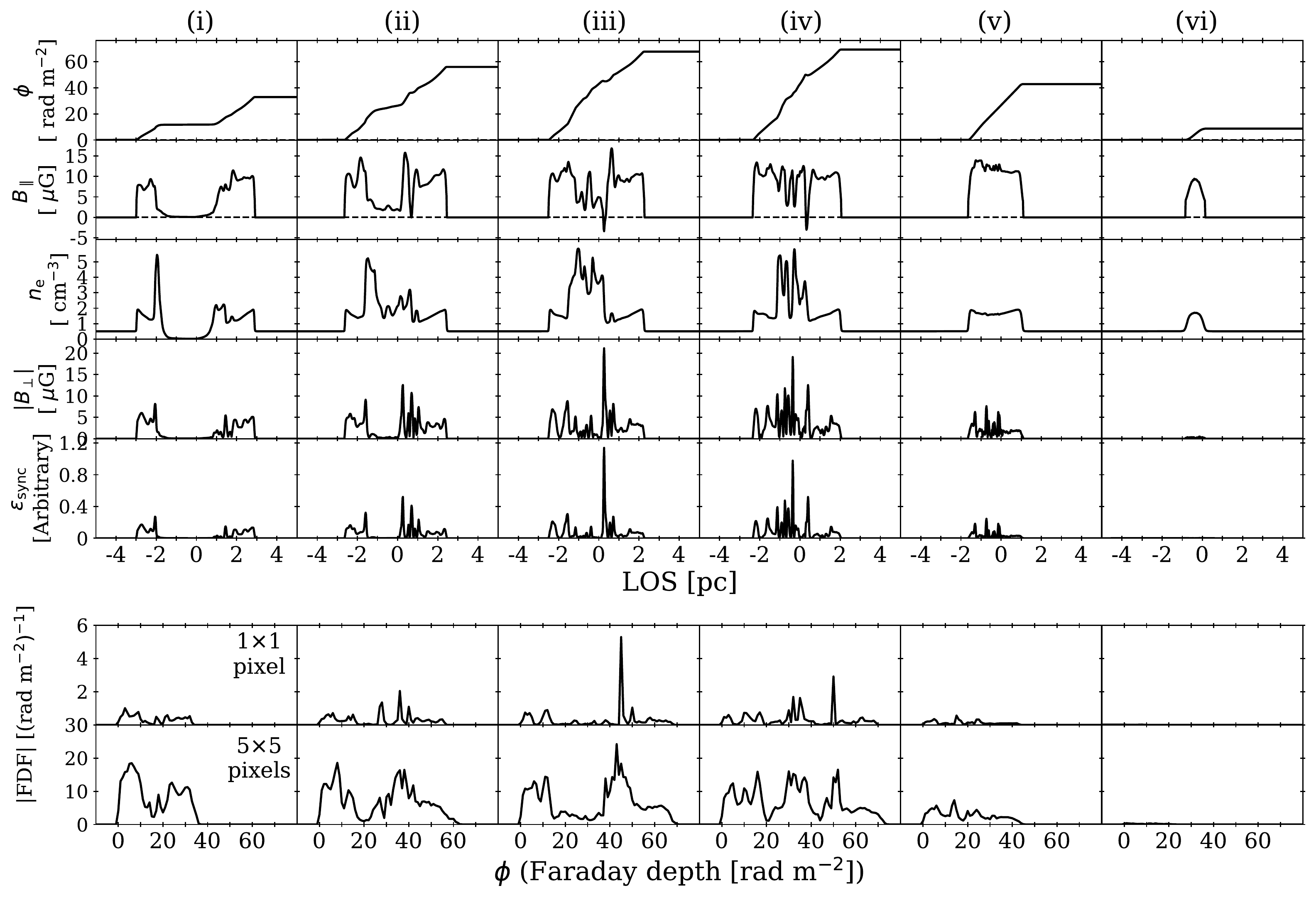}
    \caption{Same as Figure \ref{fig:fdf1}, when the viewing angle is along the $y$-axis.}
    \label{fig:fdf3}
\end{figure*}

Figures \ref{fig:fdf1} - \ref{fig:fdf3} show the profiles of five physical quantities ($\phi$, $B_{\parallel}$, $n_{\rm e}$, $B_{\perp}$ and $\varepsilon_{\rm sync}$) and the calculated FDFs along three viewing angles parallel to the $z$, $x$ and $y$-axes, respectively. As shown in Section \ref{sec:2}, the Faraday depth $\phi$ is calculated from $B_{\parallel}$ and $n_{\rm e}$, and the synchrotron emissivity $\varepsilon_{\rm sync}$ is calculated from $B_{\perp}$. Note that these plots include quantities from all 512 cells along the LOS. Finally, $\varepsilon_{\rm sync}$ and $\phi$ produce the FDFs in the sixth and seventh rows. The sixth and seventh rows show the FDFs from one pixel and multi (5$\times$5) pixels centered on the one pixel on the image plane, respectively, which corresponds to observations with different angular resolutions. The size of the Faraday depth bins is set to 0.2, 0.2 and 1.0 rad m$^{-2}$ for Figure \ref{fig:fdf1}, \ref{fig:fdf2} and \ref{fig:fdf3}, respectively, to be able to follow the distribution of the FDF properly.

The six columns in the figures display the different LOSs (i), (ii), (iii), (iv), (v) and (vi), which pass through IR, RS, SE, CD, SM and FS at the cross section sliced at the center of the SNR, respectively (see Section \ref{sec:2.1}). As mentioned in Section \ref{sec:2.1}, a single LOS can pass through more than one bubble of the turbulent pattern due to the Rayleigh-Taylor instability, and/or the two (near and far sides) shocked regions. In Figure \ref{fig:fdf1}, for example, two shocked regions can be seen in the $n_{\rm e}$ profile in the case (i): the near-side one appears at $\sim -2$ pc and far-side one at $\sim +2$ pc. In addition, many bubbles appear as spikes in the $n_{\rm e}$ profile in cases (ii), (iii) and (iv). In cases (v) and (vi), the LOSs only pass through the shocked ISM and they do not show spikes.

Figure \ref{fig:fdf1} shows the results for the case of a viewing angle along the $z$-axis which corresponds to a large $\varepsilon_{\rm sync}$ and a small $\phi$. The contribution of the background magnetic field (which is along the $y$-axis) to $B_\parallel$ is in this case zero, by definition. However, $B_{\parallel}$ is nonzero and irregular due to fluctuations and turbulence in the simulated SNR, which results in the small $\phi$ values up to $|\phi|$ $\sim$ 4 rad m$^{-2}$. The bimodal structures of $B_{\perp}$ result from the amplification of the magnetic field in the shocked ISM immediately after the contact discontinuity at the near and far sides of the shell along the LOS. The calculated FDFs (1$\times$1 pixel) have the relatively wide ($\sim 5$ rad m$^{-2}$) structures in cases (ii), (iii) and (iv), whereas they are relatively narrow and sharp in cases (v) and (vi) because polarization is concentrated in a smaller range of $\phi$.

In Figure \ref{fig:fdf2} which corresponds to the viewing angle along the $x$-axis i.e. small $\varepsilon_{\rm sync}$ and $\phi$ values, we can see that the magnetic fields are small compared with the other two cases because the initial fields are almost perpendicular to the shock front and therefore barely amplified (see also Section \ref{sec:2.1}). While $\phi$ can reach a larger value up to $|\phi|\sim$ 8 rad m$^{-2}$ at around 0~pc compared with the previous case due to the less turbulent $B_{\parallel}$, emissions are about ten times smaller. As a result, the FDFs (1$\times$1 pixel) have a slightly larger width and smaller amplitude. Again, the width gets smaller in cases (v) and (vi) because polarization is concentrated in a smaller range of $\phi$.

In Figure \ref{fig:fdf3}, since the fields almost point toward the observer, $B_{\parallel}$ is mostly positive along the LOS. Thus, $\phi$ can reach a large (positive) value, up to $\phi\sim$ 80 rad m$^{-2}$, and is an almost monotonic function of the physical distance along the LOS. Hence, the FDFs (1$\times$1 pixel) have large widths and directly reflect the distribution of the emission in physical space along the LOS. On the other hand, their amplitude is relatively small compared with the case shown in Figure \ref{fig:fdf1}, because $B_\perp$ is small. The trend for cases (v) and (vi) is the same as the other viewing angles. Although the FDF for case (vi) is barely seen due to the small emission, we can still expect from the $\phi$ distribution that its width gets smaller.

We also calculate the FDFs of multiple pixels ($5\times 5=25$ pixels) centered on the one pixel on the image plane. This corresponds to observations with five times lower angular resolution. These FDFs are shown at the bottom panels (seventh rows) of Figures \ref{fig:fdf1} - \ref{fig:fdf3}. Basically the FDFs for multiple pixels have smoother distribution compared with the FDFs for a single pixel, because a larger number of polarization cells averages out the spiky FDF pattern made by turbulent bubbles. Actually, the amplitude is not necessarily increased by a factor of 25 for 25 pixels, because the depolarization can happen on Faraday depth space when polarization from more than one computational cell are added in the same Faraday depth bin and have different polarization angles. Basically there is little change in the overall shape of the FDF and the amplitude simply increases. The exception is the case shown in Figure \ref{fig:fdf1}, where there are differences in the FDFs for the $1\times1$ and $5\times5$ pixels such as their broader shape and different locations of peaks on the Faraday depth space. This is because $B_\parallel$ is essentially coming from fluctuations and turbulence and its distribution varies more along the different pixels than for the other cases, as explained above. However, it is worth noting that the width gets smaller in cases (v) and (vi), as for the other viewing angles.

\begin{table}
 \caption{Faraday depolarization ratio.}
 \label{tab:depol}
 \begin{tabular}{ccc}
  \hline
  Frequency & VA: $z$-axis & VA: $y$-axis \\[2pt]
  GHz & ($\sigma_{\rm FDF}$ = 2.77 [rad m$^{-2}$]) & ($\sigma_{\rm FDF}$ = 19.3 [rad m$^{-2}$])\\
  \hline
  0.15 & 4.08$\times$10$^{-3}$ & 8.40$\times$10$^{-5}$\\[2pt]
  0.3 & 1.21$\times$10$^{-1}$ & 2.49$\times$10$^{-3}$\\[2pt]
  0.7 & 7.81$\times$10$^{-1}$ & 3.99$\times$10$^{-2}$\\[2pt]
  1.4 & 9.84$\times$10$^{-1}$ & 5.05$\times$10$^{-1}$\\[2pt]
  2 & 9.96$\times$10$^{-1}$ & 8.33$\times$10$^{-1}$\\
  \hline
 \end{tabular}
\end{table}

Finally, we estimate the Faraday depolarization ratio of the calculated FDFs. We choose two FDFs as representative cases: those for 5$\times$5 pixels at (iv) CD (the bottom, fourth column from the left) in Figure \ref{fig:fdf1} (VA: $z$-axis) and \ref{fig:fdf3} (VA: $y$-axis). The depolarization ratio (DP) can be calculated via
\begin{equation}
{\rm DP} = \frac{1-e^S}{S}
\end{equation}
where $S=2\sigma_{\rm FDF}^2\lambda^4$ \citep{sokoloff_etal_1998}. Note that DP = 1 means no depolarisation, while DP = 0 means total depolarisation. Here, $\sigma_{\rm FDF}$ is the RM dispersion or the width of FDF, and we define it as the standard deviation of the FDF,
\begin{equation}\label{eq:width}
\sigma_{\rm FDF} = \sqrt{\frac{\sum_j|F(\phi_j)|(\phi_j-\mu)^2}{\sum_j|F(\phi_j)|}}
\end{equation}
where
\begin{equation}
\mu = \frac{\sum_j|F(\phi_j)|\phi_j}{\sum_j|F(\phi_j)|}
\end{equation}
is the mean of the FDF. We summarize the DP at 0.15, 0.3, 0.7, 1.4, and 2.0 GHz in Table \ref{tab:depol}. With $\sigma_{\rm FDF}=2.77$ rad m$^{-2}$, the DP is about 0.4\% at 150 MHz (e.g. LOFAR band). In the case of $\sigma_{\rm FDF}=19.3$ rad m$^{-2}$, the DP is about 4\% already at 700 MHz.

\subsection{Radial profile of FDF}
\label{sec:3.2}

\begin{figure*}
	\includegraphics[width=\textwidth]{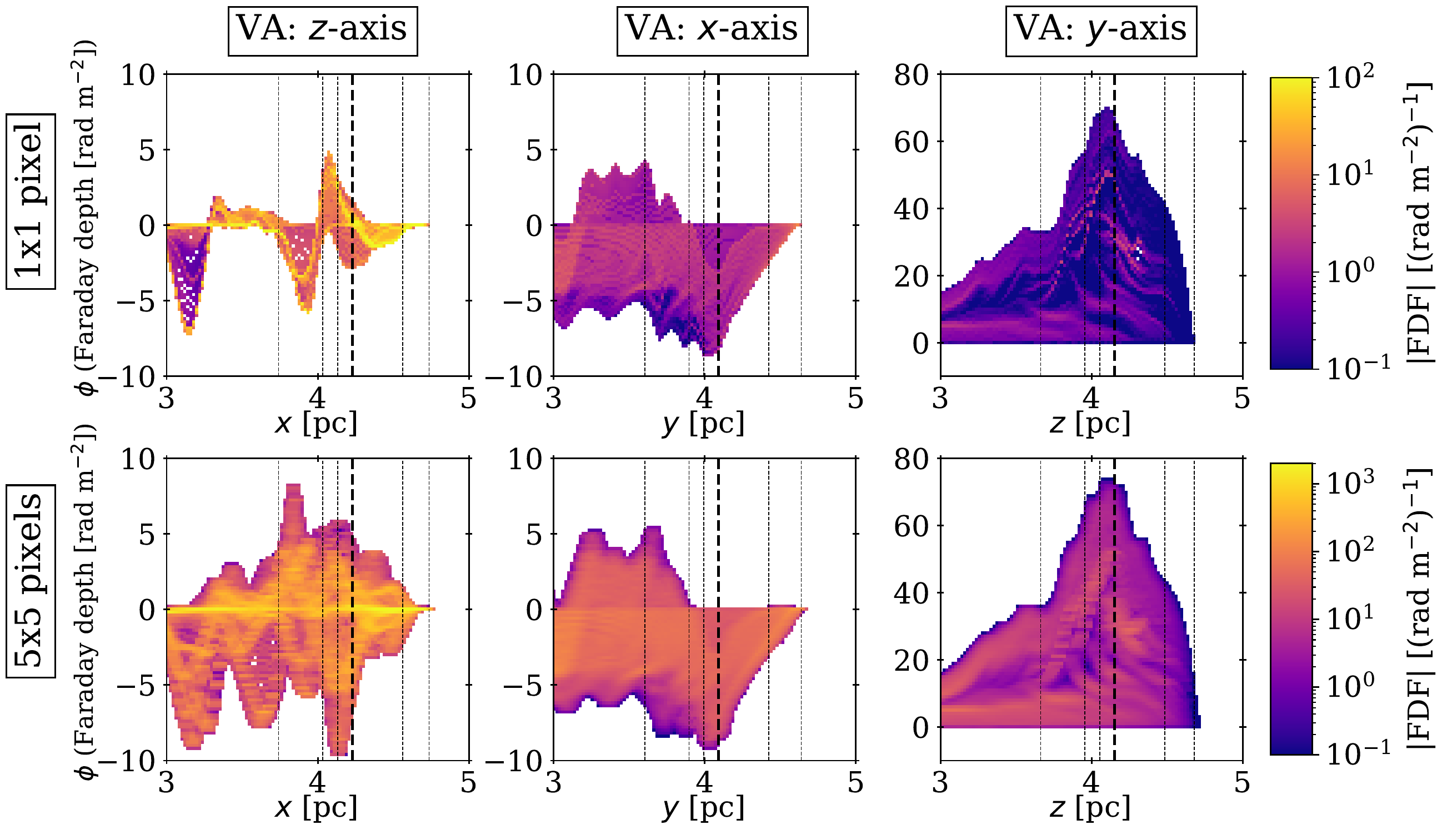}
	\caption{Radial profile of FDF. The rows and columns correspond to the number of pixels on the image plane and viewing angles (VAs), respectively. The vertical and horizontal axes are the Faraday depth and radius [pc], respectively, and the color shows the absolute amplitude of the FDF. The dashed black lines correspond to the six characteristic structures shown in Figure \ref{fig:rho}, and thick lines are for (iv).}
    \label{fig:rad_dep}
\end{figure*}

In the previous section, we have shown FDFs of the simulated SNR for the six characteristic shock structures from three different viewing angles (VAs) and found that the width of FDFs get smaller at radii larger than the (iv) CD location. In this section, we investigate the seamless radial dependence of the FDF to see how the width of FDFs changes radially in detail. Figure \ref{fig:rad_dep} shows the radial profiles of the FDFs, where the horizontal axis is the SNR radius, the vertical axis is the Faraday depth, $\phi$, and the color is the absolute amplitude of the FDF in arbitrary units. The rows and columns correspond to the number of pixels on image plane and the viewing angle, respectively. The dashed vertical lines in the plots correspond to the six characteristic shock structures, and the (iv) is plotted with the thick line.

For ``VA: $x$-axis'' and ``VA: $y$-axis'' cases, the width of FDFs get smaller outside of (iv) almost monotonically. In other words, while moving inward from the outside, (iv) is located before of the first local maximum of the width. In addition, (ii) is located after of the local maximum. Since the shock structures are defined from the radial profile of $n_{\rm e}$ at the cross section sliced at the center of the SNR sphere, the LOSs of the structures (i) - (iv) can pass through more than one bubble of the turbulent pattern due to Rayleigh-Taylor instability (see Section \ref{sec:2.1}). Thus, the size of the bubbles, which is about 0.2 - 0.3 pc as seen from Figure \ref{fig:rho}, leads to the uncertainty of the definition of the structures. This uncertainty results in the complex structure of the width of FDFs just inside the local maximum and small bumps at $z\sim 4.2$ pc in the images of ``VA: $y$-axis'' case. Nevertheless, it still holds that the width of the FDF whose LOS passes through the structure around (iv) tends to be large because the LOS passes the region where the magnetic field and thermal electron density are large for a relatively long distance. Hence, it seems to be possible to distinguish (iv) from the radial profile of the FDF width.

For ``VA: $z$-axis'' case, the profile of the FDF width outside of (iv) is rather complicated. This is because $B_\parallel$ arises from fluctuations and turbulence (see Section \ref{sec:3.1}) and the LOS distribution of $\phi$ is complicated. This may imply that it is not easy to distinguish (iv) from the radial profile of the FDF width when the regular magnetic field around the SNR is perpendicular to the LOS and to the shock normal.

\cite{warren_etal_2005} identified in X-ray observations the radius of the FS as the radius where the brightness exceeds a certain value, when moving in from the outside. Since the region outside of the (vi) has no emission coming from the SNR, the same approach can be applied to this work: the FS structure can be identified as the location where the FDF emission is below noise level.

\section{Identifying the Contact Discontinuity}
\label{sec:4}

As seen in the previous section, the width of the FDF viewed through a SNR shell tends to decrease outside the region (iv), which passes through the contact discontinuity (CD) at the cross section sliced at the center of the SNR. The reason is that the magnetic field is amplified in the shocked ISM immediately after the contact discontinuity. This leads to larger RM value and thus, the width of FDF is large. With increasing radius, the LOS no longer passes through the region where the magnetic field is amplified, and the width gets smaller. Therefore, this behavior is rather universal for SNR shocks, suggesting the prospect to find the contact discontinuity in SNR shocks from the width of FDF derived by Faraday tomography. We explore this prospect in this section.

\begin{figure}
    \begin{center}
        \includegraphics[width=\columnwidth]{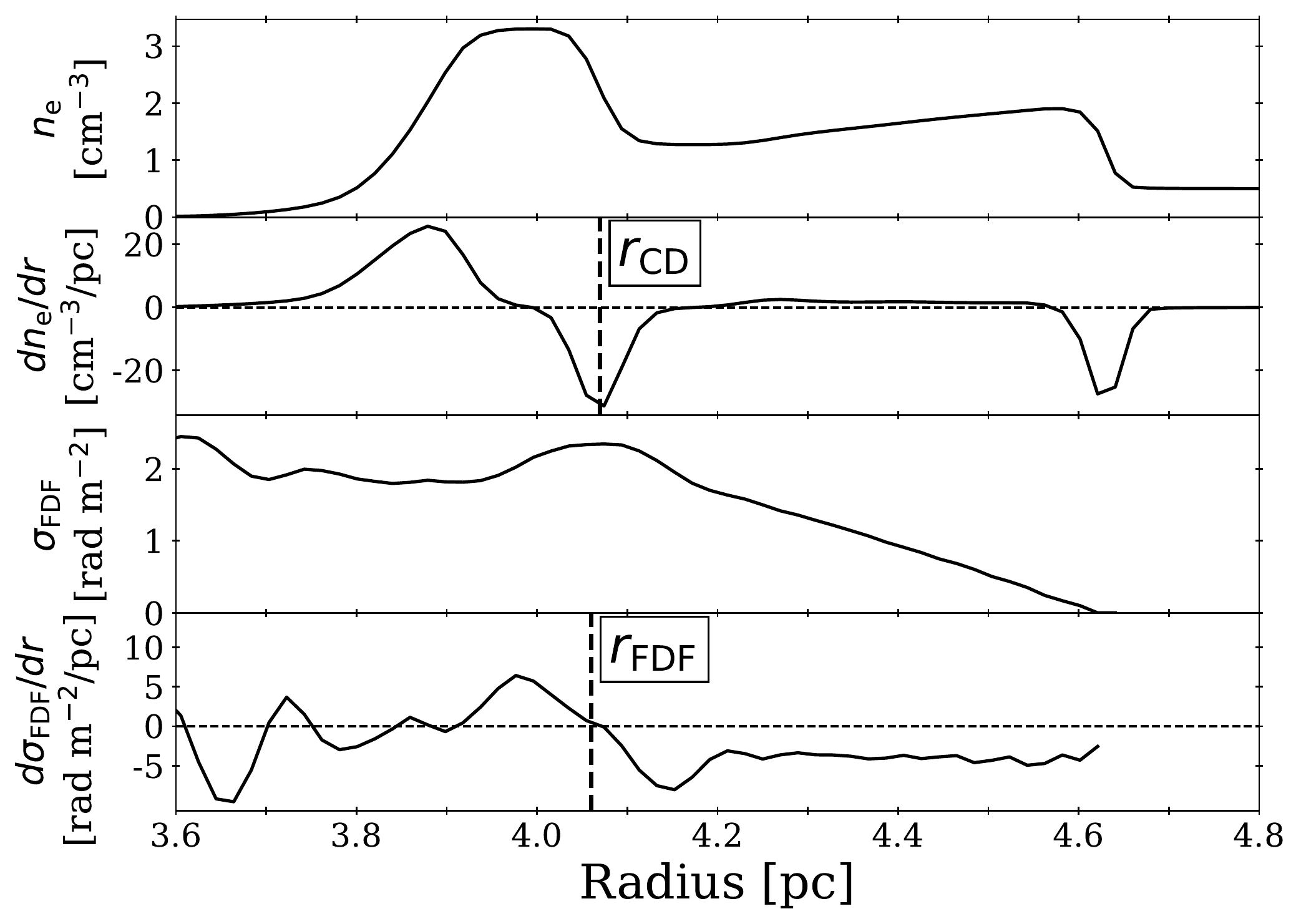}
    \end{center}
    \caption{An example of the radial profile of $n_{\rm e}$, $dn_{\rm e}/dr$, $\sigma_{\rm FDF}$ and $\sigma_{\rm FDF}/dr$ from top to bottom panels. The defined radii, $r_{\rm cd}$ and $r_{\rm FDF}$, are also put in the plot as the dashed vertical lines.}
    \label{fig:cd_def}
\end{figure}

First, we define the contact discontinuity as the radius at which the gradient of $n_{\rm e}$ is largest in its radial distribution between the locations of shocked ejecta and shocked ISM, and label it as $r_{\rm CD}$. This is shown in the top two panels in Figure \ref{fig:cd_def}. Since this is calculated at the cross section sliced at the center of the SNR, the radius can differ by the size of the turbulent bubbles if we consider the effect of geometric projection. In this work, we just regard it as the ambiguity in the definition of $r_{\rm CD}$ for simplicity. We also plot the radial dependence of the width of FDF and its gradient in the bottom two panels in Figure \ref{fig:cd_def}. Here, the width, $\sigma_{\rm FDF}$, is calculated using Equation \ref{eq:width}. As explained above, the width ($\sigma_{\rm FDF}$) gets smaller monotonically outside of the contact discontinuity. Thus, we define $r_{\rm FDF}$ as the radius where the gradient of the FDF width becomes zero, i.e., $\sigma_{\rm FDF}$ is the first local maximum while moving inward from the outside.

\begin{figure*}
    \begin{center}
	    \includegraphics[width=\textwidth]{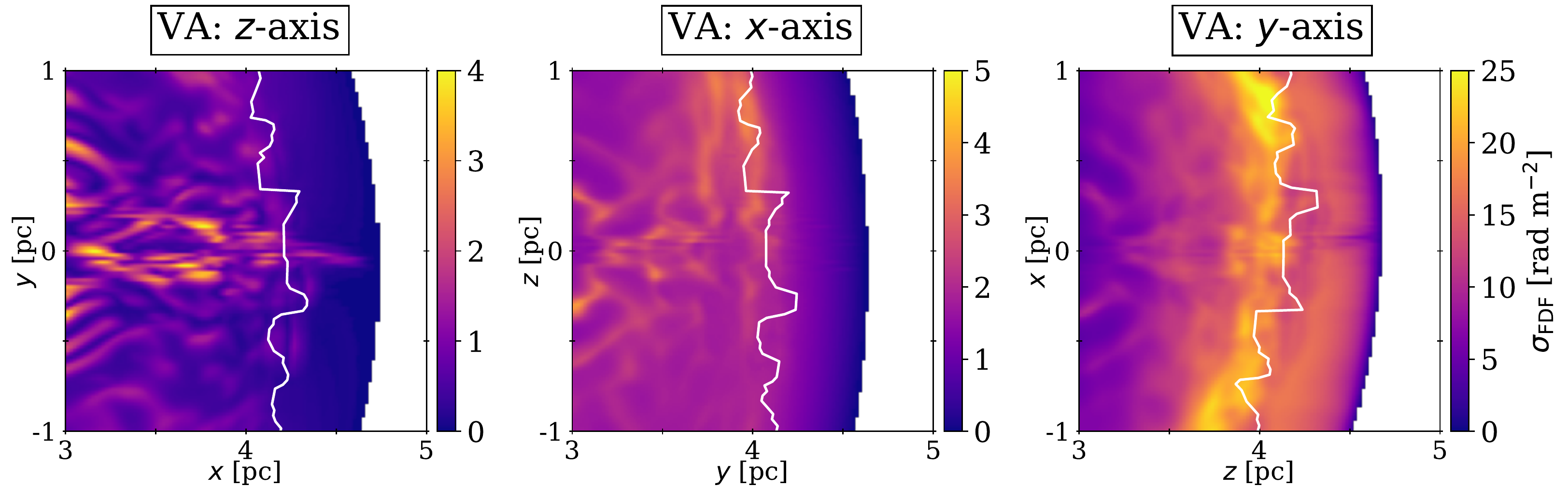}
    \end{center}

    \caption{The colormap shows the width of the FDFs, $\sigma_{\rm FDF}$, while the location of the CD, $r_{\rm CD}$, is represented by white lines. The columns correspond to different viewing angles.}
    \label{fig:cd_loc}
\end{figure*}

Figure \ref{fig:cd_loc} shows a colormap of the width of the FDFs, $\sigma_{\rm FDF}$, and the location of the CD, $r_{\rm CD}$, represented by a white line. The CD seems to coincide with the region where the FDF width starts to decrease. We calculate the $r_{\rm FDF}$ and compare it with $r_{\rm CD}$. The absolute differences between the two radii are $0.105\pm 0.089$, $0.081\pm 0.060$, and $0.171\pm 0.077$ pc for viewing angles parallel to the $z$-, $x$- and $y$- axes, respectively. These results suggest that the location of the contact discontinuity may be identified with an accuracy of 0.1 - 0.2 pc by means of Faraday tomography.

\section{Discussion}
\label{sec:5}

When we observe polarization emission from SNRs and attempt to interpret the FDFs, we need to consider several observational factors which affect the reconstruction of FDFs from observed polarization spectra. One important factor is frequency coverage, which determines the resolution in Faraday depth ($\delta\phi$) as well as the largest depth scale (max-scale) of FDF, that are expressed as
\begin{eqnarray}
&&\delta\phi\simeq\frac{2\sqrt{3}}{\Delta\lambda^2} [{\rm rad~m^{-2}}]\\
&&{\rm max\mbox{-}scale}\simeq\frac{\pi}{\lambda_{\rm min}^2} [{\rm rad~m^{-2}}]
\end{eqnarray}
\citep{brentjens_debruyn_2005}, where $\Delta\lambda^2=\lambda_{\rm max}^2-\lambda_{\rm min}^2$ and $\lambda_{\rm min}^2$ and $\lambda_{\rm max}^2$ are minimum and maximum squared wavelengths of the frequency coverage, respectively. This means that lower frequencies are sensitive to the small-scale structures of the FDF because it provides wide coverage in $\lambda^2$ space, while higher frequencies are also important to capture diffuse structures of the FDF because it is less affected by the Faraday depolarization. From Figures \ref{fig:fdf1}-\ref{fig:rad_dep}, we can see that the width of FDF is $\sim 10$ rad m$^{-2}$ when the viewing angle is along the $z$- and $x$- axes and $\sim 70$ rad m$^{-2}$ for the $y$- axis. To distinguish the difference in the width, the resolution needs to be substantially smaller than the width. In Table \ref{tab:telescopes}, we summarize $\delta\phi$ and max-scale obtained from several telescopes. We need a minimum frequency $\lesssim$ 300 MHz for FDFs with a width of 10 rad m$^{-2}$, while $\sim$ 1 GHz is sufficient for FDFs with the width of 70 rad m$^{-2}$. On the other hand, higher frequencies are also necessary to avoid Faraday depolarization (see also Table \ref{tab:depol}) and capture the whole structure of the FDF to distinguish its width. From Table \ref{tab:telescopes}, a maximum frequency $\gtrsim$ 500 MHz is enough for FDFs with 10 rad m$^{-2}$ width while $\sim$ 2 GHz is necessary for FDFs with 70 rad m$^{-2}$ width.

\begin{table}
 \caption{Specifications of radio observatories}
 \label{tab:telescopes}
 \begin{tabular}{cccc}
  \hline
  Telescope & Frequency coverage & $\delta\phi$ & max-scale\\
   & [GHz] & [rad m$^{-2}$] & [rad m$^{-2}$]\\
  \hline
  SKA1-MID Band 1 \& 2 & 0.35 - 1.76 & 4.9 & 110 \\[2pt]
  SKA1-Low & 0.05 - 0.35 & 0.098 & 4.3 \\[2pt]
  ASKAP Full band & 0.7 - 1.8 & 22 & 110 \\[2pt]
  LOFAR HBA & 0.12 - 0.17 & 1.1 & 1.0 \\[2pt]
  VLA L band & 1 - 2 & 51 & 140 \\[2pt]
  VLA L+S band & 1 - 4 & 41 & 559 \\[2pt]
  VLA L+S+C band & 1 - 7 & 39 & 1713 \\[2pt]
  VLA P band & 0.23 - 0.47 & 2.7 & 7.7 \\[2pt]
  \hline
 \end{tabular}
\end{table}

The reconstructed FDF from $P(\lambda^2)$ with limited frequency coverage has side lobes. In addition, the noise on $P(\lambda^2)$ appears as noise on the FDF at all Faraday depth. These Faraday structures, especially those at large values of |$\phi$|, strongly affect the estimation of the width of the FDF. For instance, \cite{dickey_etal_2019} used Global Magneto-Ionic Medium Survey (GMIMS) data to study the three dimensional magnetic field structure of the Milky Way using zeroth, first, and second moments of the FDF. To avoid the issue mentioned above, they set a threshold on the FDF, and calculate the moments around the peak until the spectrum drops below the threshold. This method could be useful for an accurate estimation of the width of FDFs reconstructed from the data with limited frequency coverage.

Another important observational factor that we briefly addressed is the angular resolution. We see that the appearance of FDF significantly depends on the angular resolution. As the resolution gets worth, the so-called beam depolarization effect becomes serious and we lose polarized emission. In this situation, it would be difficult to distinguish the real flux from noise and thus, the estimation of the FDF width would also be difficult. In addition, if the SNRs are not well resolved because the angular resolution is low or because they are at large distance, the radial profile of the FDF structure would also not be well resolved and it would be difficult to distinguish the width of the FDF.

Our simulation clearly showed that the viewing angle is one of the key parameter for observable FDF; the resultant FDF primarily depends on the angle between the initial regular magnetic field and the line-of-sight. On the one hand, such a dependence is useful to study the magnetic field orientation, but, on the other hand, it implies that our interpretation of the FDF will suffer from an uncertainty on the angle. For instance, \cite{west_etal_2016} reproduce the observed morphologies of SNRs through a simple and physically motivated model of an SNR expanding into a regular magnetic field. If the large-scale Galactic magnetic field (GMF) model is known, their method can predict the SNR distance by comparing simulated and observed bilateral axis angle of SNRs. Conversely, if SNR distances are known, the method can constrain the large-scale GMF at their position. Their method could resolve the uncertainty on the angle in our method, and once the uncertainty is resolved, the two methods are complementary to constrain the regular magnetic field orientation around the SNR positions. Note that the identification of the contact discontinuity using the width of the FDF is less affected by the viewing angle as shown in Section \ref{sec:4}.

In this work, we made the simple assumption that the cosmic-ray electron (CRe) density is constant. However, the mechanism of CRe injection is still in debate. In the particular case of SN1006, three scenarios are often considered: electrons mainly distributed in the quasi-perpendicular regions of the shock \citep[favored by e.g.,][]{petruk_etal_2009,schneiter_etal_2010}, electrons mainly distributed in quasi-parallel regions of the shock \citep[e.g.,][]{bocchino_etal_2011,schneiter_etal_2015,winner_etal_2020}, and electrons uniformly distributed along the shock \citep[e.g.,][]{schneiter_etal_2010,west_etal_2016} (Note that our assumption corresponds to the latter case). The assumption of a constant spectral index of synchrotron radiation is another simplification of the model. The strong magnetic field (bottom panel of Figure~\ref{fig:mag_3d}) leads to significant synchrotron loss of CRe and hence to spectral steepening. Nevertheless, they have little impact to the conclusion that the contact discontinuity may be identified from the FDF width, since they only affect the amplitude of the FDF and not the Faraday depth.

In our simulation of SNR, we considered that the magnetic field in the ambient ISM is initially regular, whereas we generally expect it to be turbulent. Moreover, we considered a uniform, fully-ionized plasma as the ambient ISM. In reality, it is likely that the ambient plasma is disturbed and possesses density fluctuations. Such fluctuations are strongly amplified and can play an important role for the evolution of radio SNRs. We also ignored other mechanisms of turbulent field amplification to keep the interpretation of the results simple. To examine the effects of such strong turbulent magnetic field, MHD simulations with the density fluctuations in the ISM are required. Nevertheless, we note that the FDF can have information of magnitude and scale of turbulent magnetic field along the line of sight as the degree of Faraday dispersion \citep{ideguchi_etal_2017}. We will study this in the forthcoming paper. On the other hand, the magnetic field amplified at the upstream region due to Bell amplification \citep{bell_etal_2004} may be efficiently damped at the downstream \citep{pohl_etal_2005}. It would be useful to compare future SNR observations with our results based on these simple assumptions. The discrepancies revealed by such a comparison could provide information on the complex nature of the ambient ISM, which is for instance important to understand particle acceleration in SNRs.

\section{Conclusion}
\label{sec:6}

We predicted the FDF of an SNR for the first time using 3D MHD simulations. The FDF shows the complex profile including peaked, Faraday-thin components and diffuse, Faraday-thick components. This suggests that the Faraday tomography is necessary for a detailed study of the magnetic field in SNRs from radio polarization observations, especially thanks to its potential to solve the difficulty of separating foreground and internal RM contributions. Furthermore, the wide-band polarimetry of the new generation radio telescopes is important to reconstruct the high-quality FDF. This prediction would be useful to interpret FDFs reconstructed from real observations, to discriminate the real Faraday structure of SNRs from ambiguities such as so-called RM ambiguity \citep{farnsworth_etal_2011,kumazaki_etal_2014,miyashita_etal_2016}, and for the appropriate functions for the QU-fitting \citep{sokoloff_etal_1998,osullivan_etal_2012,ideguchi_etal_2014,miyashita_etal_2019}. Although the FDF depends on the angle between the LOS and the regular ambient magnetic field (viewing angle), this degeneracy may be resolved by estimating the field orientation from the global morphology of the SNRs \citep{west_etal_2016}.

We also focused on the radial dependence of the width of the FDF and found that it decreases monotonically outside of the contact discontinuity (CD). Using this phenomenon, we demonstrated that the location of the CD can be identified with an accuracy of 0.1-0.2 pc. Furthermore, we found that the accuracy does not depend on the viewing angle. The identification of the CD location in the SNR shock using radio polarimetry by means of the Faraday tomography technique would be a complementary approach with that using X-ray observations and hence, the technique would also be important for the study of the particle acceleration in SNRs.

\section*{Acknowledgements}

We thank the anonymous referee for constructive comments and suggestions. SI thanks Marijke Haverkorn for fruitful discussions and suggestions on this work. This work is part of the joint NWO-CAS research programme in the field of radio astronomy with project number 629.001.022, which is (partly) financed by the Dutch Research Council (NWO). TA is supported in part by JSPS Grant-in-Aid for Scientific Research 21H01135. KT is partially supported by JSPS KAKENHI Grant Numbers 15H05896, 16H05999, 17H01110, 20H00180, 21H01130, and 21H04467, Bilateral Joint Research Projects of JSPS, and the ISM Cooperative Research Program (2020-ISMCRP-2017, 2021-ISMCRP-2017).

\section*{Data availability}

The data underlying this article will be shared on reasonable request to the corresponding author.

\bibliographystyle{mnras}
\bibliography{main}

\appendix

\section{Maps for different aspect angles}
\label{sec:app1}

Here, we compare our results with that of \cite{kothes_brown_2009} (hereafter KB09). The two works share the similar simple assumption in which the SNR expands into an ambient ISM of constant gas density and regular magnetic field. Our model solves the expansion numerically which give rise to more realistic phenomena such as Rayleigh-Taylor instability than the analytical model of KB09.

\begin{figure*}
	\includegraphics[width=1.0\textwidth]{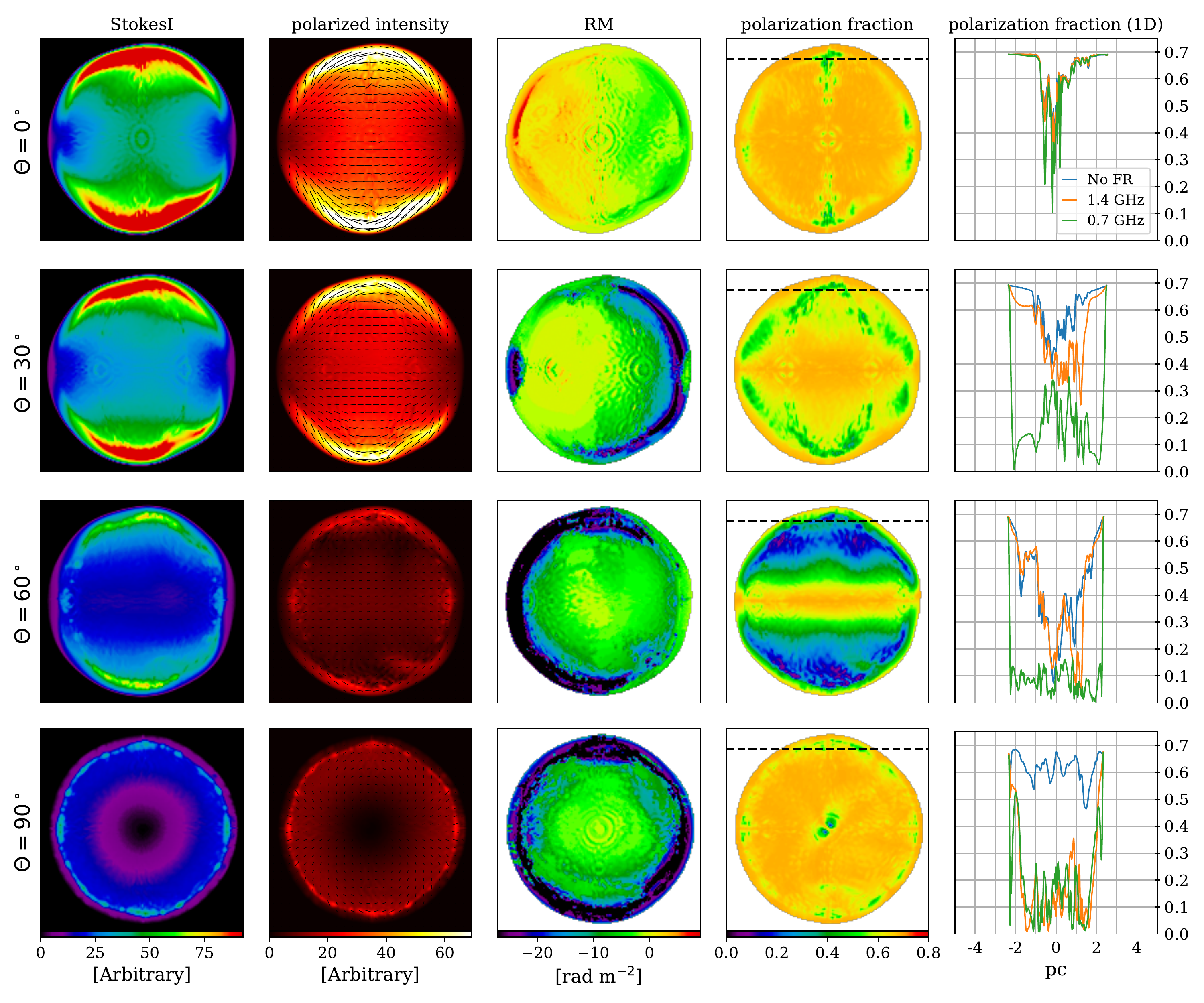}
	\caption{(Columns) Simulated synchrotron maps in total and linearly polarized intensities with overlaid magnetic field orientations, RM maps, linear polarization fraction maps, and 1D profile of the fraction along the dashed black line in the fraction maps, from left to right. Note that the Faraday rotation is not considered in these maps. In the 1D profiles, we plot the fraction taking into account the Faraday rotation at 1.4 (orange) and 0.7 (green) GHz, and without Faraday rotation (blue lines with label ``No FR''). (Rows) Different aspect angles $\Theta$ = 0$^\circ$, 30$^\circ$, 60$^\circ$, 90$^\circ$, from top to bottom.}
    \label{fig:kb09}
\end{figure*}

KB09 simulated emission structures in Stokes I, polarized intensity, and rotation measure (RM), and predicted that RM should reveal a gradient along both shells. In Figure \ref{fig:kb09}, we show the projected synchrotron maps in total and polarized intensities and the RM maps for different aspect angles $\Theta$ to allow comparison with Figure 2 of KB09. The magnetic field orientations (the black bars in the polarized intensity maps) show the polarization planes rotated by 90$^\circ$. To compute these maps, polarized radiative transfer was performed without accounting for Faraday rotation. To simulate the RM maps, we calculate the FDF of each pixel and pick up the Faraday depth of its peak as the RM values. This would be equivalent to the RM which is estimated by linear fit of $\chi$ vs. $\lambda^2$ at the high enough frequency where the Faraday structure of SNRs is seen as Faraday simple.

Our results basically agree to those of KB09. The small-scale fluctuations in our maps arise from fluctuations in $n_{\rm e}$ and magnetic fields due to the Rayleigh-Taylor instability. There is also an obvious difference in the $\Theta$ = 60$^\circ$ RM map: the KB09 map shows a large structure with peaking RM that we do not observe. Since the structure in KB09 has positive RM, it corresponds to a region where the fields point towards us. Given that the aspect angle $\Theta$ is 60$^\circ$, such fields are distributed at the downstream of the far-side shell along the line of sight (LOS). Thus only emission from regions further away than the downstream region along the LOS (e.g. the upstream region) can have positive RM. Since we ignore the upstream fields and the dominant emission comes from the downstream in our model (see also bottom panel of Figure \ref{fig:mag_3d} and Figures~\ref{fig:fdf1}-\ref{fig:fdf3}), the emission has hardly any positive RM. On the other hand, although there is no detailed explanation about the synchrotron emission distribution in KB09 model, we suspect that the dominant emission in their model comes from the upstream region and thus, it has positive RM. Hence the extra structure in their map.

In addition to the three kinds of maps mentioned above, we also show polarization fraction maps and their 1D profiles taking into account the Faraday rotation at 1.4 (orange) and 0.7 (green) GHz, and without Faraday rotation (blue lines with label ``No FR''). When $\Theta=0^\circ$, the perpendicular field along the LOS is dominant and regular at the bright rim and the emission is less affected by Faraday depolarization and wavelength-independent depolarization. In the case of $0^\circ<\Theta<90^\circ$, both of the two types of depolarization occur at the bright rim because the LOS fields cause Faraday rotation, and because the projected-on-sky fields at near and far side shells along the LOS have different orientations. When $\Theta=90^\circ$, wavelength-independent depolarization does not occur because the projected-on-sky fields have the same orientation at near and far side, while the amount of Faraday depolarization is large because the LOS fields are dominant. The 1D profiles also show the frequency dependence on the polarization fraction. When depolarization is effective, the degree of polarization is smaller at lower frequency and ranges from a few to a few tens of percent.

\bsp	
\label{lastpage}
\end{document}